\documentclass[conference,a4paper]{IEEEtran}
\IEEEoverridecommandlockouts
\usepackage{cite}
\usepackage{amsmath,amssymb,amsfonts,mathrsfs}
\usepackage{bm}
\usepackage{algorithm}
\usepackage{algpseudocode}
\usepackage{graphicx}
\usepackage{textcomp}
\usepackage{xcolor}
\usepackage[nolist]{acronym} 
\usepackage{pgfplots}
\pgfplotsset{compat=1.18}
\usepackage{url}
\usepackage[caption=false,font=footnotesize]{subfig} 
\usepackage{siunitx}
\RequirePackage{amsmath,amssymb,amsfonts,mathrsfs,bm,xspace,upgreek}
\newcommand{\transpose}{\top}

\newcommand{\safemath}[2]{\newcommand{#1}{\ensuremath{#2}\xspace}}

\newcommand{\mathlabel}[1]{\mathrm{#1}}

\makeatletter
\newcommand{\notation@alphabet}[3]{%
  \def\notation@define##1{%
    \expandafter\safemath\csname #1##1\endcsname{#2{##1}}}%
  \@tfor\notation@letter:=#3\do{%
    \expandafter\notation@define\notation@letter}}
\newcommand{\notation@lower}[2]{%
  \notation@alphabet{#1}{#2}{abcdefghijklmnopqrstuvwxyz}}
\newcommand{\notation@upper}[2]{%
  \notation@alphabet{#1}{#2}{ABCDEFGHIJKLMNOPQRSTUVWXYZ}}
\newcommand{\notation@plain}[1]{#1}
\newcommand{\notation@vector}[1]{\csname vec#1\endcsname}
\newcommand{\notation@matrix}[1]{\csname mat#1\endcsname}

\notation@upper{u}{\mathrm}
\notation@lower{bm}{\mathbf}
\notation@upper{b}{\mathbf}
\notation@lower{bmi}{\bm}
\notation@upper{bim}{\bm}

\notation@upper{set}{\mathcal}
\notation@upper{col}{\mathscr}
\notation@upper{op}{\mathbb}

\notation@lower{vec}{\bm}
\notation@upper{mat}{\pmb}

\notation@lower{rnd}{\notation@plain}
\notation@lower{rvec}{\notation@vector}
\notation@upper{rmat}{\notation@matrix}
\makeatother

\safemath{\emptySet}{\varnothing} 
\safemath{\identityop}{\opI}      
\safemath{\opZero}{\mathbb{O}}    
\safemath{\bmzero}{\mathbf{0}}    
\safemath{\bmone}{\mathbf{1}}     
\safemath{\bZero}{\bmzero}        
\safemath{\bOne}{\bmone}          
\safemath{\veczero}{\bmzero}      
\safemath{\vecone}{\bmone}        
\safemath{\matzero}{\bmzero}      
\safemath{\matone}{\bmone}        
\safemath{\matidentity}{\matI}    

\safemath{\bmalpha}{\bm{\alpha}}   
\safemath{\bmxi}{\bm{\xi}}         
\safemath{\bmlambda}{\bm{\lambda}} 
\safemath{\bmmu}{\bm{\mu}}         
\safemath{\bmtheta}{\bm{\theta}}   
\safemath{\bmphi}{\bm{\phi}}       
\safemath{\bmgamma}{\bm{\gamma}}   
\safemath{\bmgama}{\bmgamma}       

\safemath{\vecalpha}{\bmalpha}      
\safemath{\vecxi}{\pmb{\xi}}        
\safemath{\veclambda}{\pmb{\lambda}}
\safemath{\vecmu}{\pmb{\mu}}        
\safemath{\vectheta}{\pmb{\theta}}  
\safemath{\vecphi}{\pmb{\phi}}      
\safemath{\vecgamma}{\pmb{\gamma}}  
\safemath{\vecrho}{\pmb{\rho}}      
\safemath{\vecell}{\pmb{\ell}}      
\safemath{\vecomega}{\pmb{\omega}}  
\safemath{\vecDelta}{\pmb{\Delta}}  

\safemath{\rvecxi}{\bmxi}          
\safemath{\rveclambda}{\bmlambda}  
\safemath{\rvecmu}{\bmmu}          
\safemath{\rvectheta}{\bmtheta}    
\safemath{\rvecphi}{\bmphi}        
\safemath{\rvecgamma}{\bmgamma}    

\safemath{\bDelta}{\mathbf{\Delta}}    
\safemath{\bLambda}{\mathbf{\Uplambda}}
\safemath{\bPhi}{\mathbf{\Upphi}}      
\safemath{\bSigma}{\mathbf{\Upsigma}}  
\safemath{\bOmega}{\mathbf{\Upomega}}  
\safemath{\bTheta}{\mathbf{\Uptheta}}  

\safemath{\matDelta}{\bDelta}   
\safemath{\matLambda}{\bLambda} 
\safemath{\matPhi}{\bPhi}       
\safemath{\matSigma}{\bSigma}   
\safemath{\matOmega}{\bOmega}   
\safemath{\matTheta}{\bTheta}   

\safemath{\bimDelta}{\bm{\Delta}}   
\safemath{\bimLambda}{\bm{\Lambda}} 
\safemath{\bimPhi}{\bm{\Phi}}       
\safemath{\bimSigma}{\bm{\Sigma}}   
\safemath{\bimOmega}{\bm{\Omega}}   
\safemath{\bimTheta}{\bm{\Theta}}   

\safemath{\rmatDelta}{\bimDelta}   
\safemath{\rmatLambda}{\bimLambda} 
\safemath{\rmatPhi}{\bimPhi}       
\safemath{\rmatSigma}{\bimSigma}   
\safemath{\rmatOmega}{\bimOmega}   
\safemath{\rmatTheta}{\bimTheta}   

\DeclareMathOperator{\argmin}{arg\,min} 
\DeclareMathOperator{\argmax}{arg\,max} 

\newtheorem{theorem}{Theorem}

\algrenewcommand\alglinenumber[1]{\scriptsize #1:}
\begin{acronym}
 \acro{E2E}{end-to-end}
 \acro{CSI}{channel state information}
 \acro{RCU}{random-coding union}
 \acro{RCUs}{random-coding union bound with parameter $s$}
 \acro{SNR}{signal-to-noise ratio}
 \acro{TurboAE}{turbo autoencoder}
 \acro{PSK}{phase-shift keying}
 \acro{NR}{5G New Radio}
 \acro{CRC11}{11-bit cyclic redundancy check}
 \acro{SCL}{successive cancellation list}
 \acro{FCN}{fully connected network}
 \acro{FNN}{feedforward neural network}
 \acro{UE}{user equipment}
 \acro{UL}{uplink}
 \acro{BS}{basestation}
 \acro{TDD}{time division duplex}
 \acro{FDD}{frequency division duplex}
 \acro{ECC}{error-correcting code}
 \acro{HDD}{hard decision decoding}
 \acro{SDD}{soft decision decoding}
 \acro{NND}{neural network decoding}
 \acro{CNN}{convolutional neural network}
 \acro{ML}{maximum likelihood}
 \acro{GPU}{graphical processing unit}
 \acro{BP}{belief propagation}
 \acro{LTE}{long-term evolution}
 \acro{BER}{bit error rate}
 \acro{DER}{detection error rate}
 \acro{ReLU}{rectified linear unit}
 \acro{BPSK}{binary phase shift keying}
 \acro{QPSK}{quadrature phase shift keying}
 \acro{AWGN}{additive white Gaussian noise}
 \acro{MSE}{mean squared error}
 \acro{LLR}{log-likelihood ratio}
 \acro{MAP}{maximum a posteriori}
 \acro{BCE}{binary cross-entropy}
 \acro{CE}{cross-entropy}
 \acro{BLER}{block error rate}
 \acro{SQR}{signal-to-quantisation-noise-ratio}
 \acro{MIMO}{multiple-input multiple-output}
 \acro{OFDM}{orthogonal frequency division multiplex}
 \acro{RF}{radio frequency}
 \acro{LOS}{line of sight}
 \acro{NLoS}{non-line of sight}
 \acro{NMSE}{normalized mean squared error}
 \acro{CFO}{carrier frequency offset}
 \acro{SFO}{sampling frequency offset}
 \acro{IPS}{indoor positioning system}
 \acro{MIMO}{multiple-input multiple-output}
 \acro{NN}{neural network}
 \acro{LoS}{line-of-sight}
 \acro{CDF}{cumulative distribution function}
 \acro{DeepL}{deep-learning}
 \acro{DL}{deep learning}
 \acro{SISO}{single-input single-output}
 \acro{SNN}{scaled nearest-neighbor}
 \acro{SGD}{stochastic gradient descent}
 \acro{CP}{cyclic prefix}
 \acro{LMMSE}{linear minimum mean square error}
 \acro{FCL}{fully connected layer}
 \acro{DNN}{dense neural network}
 \acro{FIR}{finite impulse response}
 \acro{ELU}{exponential linear unit}
 \acro{LDPC}{low-density parity-check}
 \acro{IDD}{iterative demapping and decoding}
 \acro{IEDD}{iterative equalization, demapping and decoding}
 \acro{MMSE}{minimum mean square error}
 \acro{QAM}{quadrature amplitude modulation}
 \acro{URLLC}{ultra-reliable low-latency communication}
 \acro{FEC}{forward error correction}
 \acro{MTC}{machine-type communications}
 \acro{mMTC}{massive machine-type communications}
 \acro{STO}{sampling time offset}
 \acro{STE}{straight-through estimator}
 \acro{CCE}{categorical cross-entropy}
 \acro{IoT}{Internet of Things}
 \acro{CRC}{cyclic redundancy check}
 \acro{MD}{missed detection}
 \acro{FA}{false alarm}
 \acro{FAR}{false alarm rate}

\end{acronym}

\definecolor{mittelblau}{rgb}{0.000000,0.494118,0.776471}
\definecolor{black}{rgb}{0.000902,0.000863,0.000863}
\definecolor{rot}{rgb}{0.94,0.03,0.06}
\definecolor{apfelgruen}{rgb}{0.549020,0.776471,0.243137}
\definecolor{gelb}{rgb}{0.003922,0.866667,0.000000}
\definecolor{orange}{rgb}{0.956863,0.435294,0.129412}
\definecolor{pink}{rgb}{0.929412,0.000000,0.549020}
\definecolor{lila}{rgb}{0.501961,0.039216,0.568627}
\definecolor{hellgrau}{rgb}{0.878431,0.878431,0.878431}
\definecolor{mittelgrau}{rgb}{0.501961,0.501961,0.501961}
\definecolor{dunkelgrau}{rgb}{0.313725,0.313725,0.313725}
\definecolor{anthrazit}{rgb}{0.074510,0.121569,0.121569}
\definecolor{darkgreen}{rgb}{0.125,0.5,0.169}
\definecolor{platinum}{rgb}{0.949020,0.952941,0.956863}
\definecolor{yellow}{rgb}{1.00,0.70,0.00}
\definecolor{purple}{rgb}{0.54117647,0.42745098,0.76862745}
\definecolor{gray}{rgb}{0.66,0.68,0.69}

\definecolor{e2ePurple}{rgb}{0.392157,0.270588,0.745098}
\definecolor{e2ePurpleFill}{rgb}{0.968627,0.952941,1.000000}
\definecolor{e2eGreen}{rgb}{0.137255,0.490196,0.215686}
\definecolor{e2eGreenFill}{rgb}{0.945098,0.984314,0.945098}
\definecolor{e2eBlue}{rgb}{0.086275,0.403922,0.823529}
\definecolor{e2eBlueFill}{rgb}{0.945098,0.968627,1.000000}
\definecolor{e2eGold}{rgb}{0.705882,0.478431,0.070588}
\definecolor{e2eGoldFill}{rgb}{1.000000,0.980392,0.921569}
\definecolor{e2eRed}{rgb}{0.960784,0.215686,0.215686}
\definecolor{e2eRedFill}{rgb}{1.000000,0.949020,0.949020}
\definecolor{e2eDark}{rgb}{0.200000,0.254902,0.360784}
\definecolor{e2eDrySage}{rgb}{0.639216,0.694118,0.541176}
\definecolor{e2eDrySageFill}{rgb}{0.913725,0.929412,0.788235}
\definecolor{e2eGrayFill}{rgb}{0.913725,0.925490,0.937255}
\definecolor{e2eGray}{rgb}{0.807843,0.831373,0.854902}

\pgfplotscreateplotcyclelist{corporate colours markers}{%
rot, every mark/.append style={fill=.!80!rot},mark=*\\%
mittelblau, every mark/.append style={fill=.!80!mittelblau},mark=square*\\%
apfelgruen, every mark/.append style={fill=.!80!apfelgruen},mark=triangle*\\%
orange, mark=star\\%
pink, every mark/.append style={fill=.!80!pink},mark=diamond*\\%
violettblau, every mark/.append style={fill=.!80!violettblau},mark=otimes*\\%
lila, mark=|\\%
gelb, every mark/.append style={fill=.!80!gelb},mark=pentagon*\\%
hellgrau, mark=text,text mark=p\\%
anthrazit, mark=text,text mark=a\\%
}

\pgfplotscreateplotcyclelist{corporate colours markers double}{%
rot, every mark/.append style={fill=.!80!rot},mark=*\\%
rot, dashed, every mark/.append style={fill=.!80!rot,solid},mark=*\\%
mittelblau, every mark/.append style={fill=.!80!mittelblau},mark=square*\\%
mittelblau, dashed, every mark/.append style={fill=.!80!mittelblau,solid},mark=square*\\%
apfelgruen, every mark/.append style={fill=.!80!apfelgruen},mark=triangle*\\%
apfelgruen, dashed, every mark/.append style={fill=.!80!apfelgruen,solid},mark=triangle*\\%
orange, mark=star\\%
orange, dashed, mark=star\\%
pink, every mark/.append style={fill=.!80!pink},mark=diamond*\\%
pink, dashed, every mark/.append style={fill=.!80!pink,solid},mark=diamond*\\%
violettblau, every mark/.append style={fill=.!80!violettblau},mark=otimes*\\%
violettblau, dashed, every mark/.append style={fill=.!80!violettblau,solid},mark=otimes*\\%
lila, mark=|\\%
lila, dashed, mark=|\\%
gelb, every mark/.append style={fill=.!80!gelb},mark=pentagon*\\%
gelb, dashed, every mark/.append style={fill=.!80!gelb,solid},mark=pentagon*\\%
}

\pgfplotsset{
  colormap/magma/.style={%
    /pgfplots/colormap={magma}{%
      rgb=(0.001462, 0.000466, 0.013866)
      rgb=(0.035520, 0.028397, 0.125209)
      rgb=(0.102815, 0.063010, 0.257854)
      rgb=(0.191460, 0.064818, 0.396152)
      rgb=(0.291366, 0.064553, 0.475462)
      rgb=(0.384299, 0.097855, 0.501002)
      rgb=(0.475780, 0.134577, 0.507921)
      rgb=(0.569172, 0.167454, 0.504105)
      rgb=(0.664915, 0.198075, 0.488836)
      rgb=(0.761077, 0.231214, 0.460162)
      rgb=(0.852126, 0.276106, 0.418573)
      rgb=(0.925937, 0.346844, 0.374959)
      rgb=(0.969680, 0.446936, 0.360311)
      rgb=(0.989363, 0.557873, 0.391671)
      rgb=(0.996580, 0.668256, 0.456192)
      rgb=(0.996727, 0.776795, 0.541039)
      rgb=(0.992440, 0.884330, 0.640099)
      rgb=(0.987053, 0.991438, 0.749504)
    },
  },
  colormap/inferno/.style={%
    /pgfplots/colormap={inferno}{%
      rgb=(0.001462, 0.000466, 0.013866)
      rgb=(0.037668, 0.025921, 0.132232)
      rgb=(0.116656, 0.047574, 0.272321)
      rgb=(0.217949, 0.036615, 0.383522)
      rgb=(0.316282, 0.053490, 0.425116)
      rgb=(0.410113, 0.087896, 0.433098)
      rgb=(0.503493, 0.121575, 0.423356)
      rgb=(0.596940, 0.154848, 0.398125)
      rgb=(0.688653, 0.192239, 0.357603)
      rgb=(0.775059, 0.239667, 0.303526)
      rgb=(0.851384, 0.302260, 0.239636)
      rgb=(0.912966, 0.381636, 0.169755)
      rgb=(0.956852, 0.475356, 0.094695)
      rgb=(0.981895, 0.579392, 0.026250)
      rgb=(0.987464, 0.690366, 0.079990)
      rgb=(0.973088, 0.805409, 0.216877)
      rgb=(0.947594, 0.917399, 0.410665)
      rgb=(0.988362, 0.998364, 0.644924)
    },
  },
  colormap/plasma/.style={%
    /pgfplots/colormap={plasma}{%
      rgb=(0.050383, 0.029803, 0.527975)
      rgb=(0.186213, 0.018803, 0.587228)
      rgb=(0.287076, 0.010855, 0.627295)
      rgb=(0.381047, 0.001814, 0.653068)
      rgb=(0.471457, 0.005678, 0.659897)
      rgb=(0.557243, 0.047331, 0.643443)
      rgb=(0.636008, 0.112092, 0.605205)
      rgb=(0.706178, 0.178437, 0.553657)
      rgb=(0.768090, 0.244817, 0.498465)
      rgb=(0.823132, 0.311261, 0.444806)
      rgb=(0.872303, 0.378774, 0.393355)
      rgb=(0.915471, 0.448807, 0.342890)
      rgb=(0.951344, 0.522850, 0.292275)
      rgb=(0.977856, 0.602051, 0.241387)
      rgb=(0.992541, 0.687030, 0.192170)
      rgb=(0.992505, 0.777967, 0.152855)
      rgb=(0.974443, 0.874622, 0.144061)
      rgb=(0.940015, 0.975158, 0.131326)
    },
  },
}

\def\BibTeX{{\rm B\kern-.05em{\sc i\kern-.025em b}\kern-.08em
    T\kern-.1667em\lower.7ex\hbox{E}\kern-.125emX}}
\begin{document}

\title{Learning Joint Synchronization, Channel Estimation, and Decoding: Comparison with a Finite-Blocklength Achievability Bound}



\author{\IEEEauthorblockN{Bingcheng Chen, Giuseppe Durisi, Christian Häger}
\IEEEauthorblockA{
Department of Electrical Engineering, Chalmers University of Technology, 41296 Gothenburg, Sweden\\ 
}
\thanks{This work was supported in part by the Swedish Foundation for Strategic Research (SSF), under grant FUS21-0004. The computation resources were partly provided by the National Academic Infrastructure for Supercomputing in Sweden (NAISS), partially funded by the Swedish Research Council through grant agreement no. 2022-06725.}
}

\maketitle

\begin{abstract}
%
Learned schemes that jointly perform tasks such as synchronization and decoding have achieved promising results for short-packet communication. However, it remains unclear how close their performance is to information-theoretic bounds. 
To investigate this question, we extend the \ac{TurboAE} proposed in D{\"o}rner et al. (2023) to a single-input single-output waveform channel with an unknown delay and memoryless block fading, for which an achievability bound from finite-blocklength information theory is available. 
Our extension incorporates a pilot-free, model-guided neural synchronizer and enables joint synchronization, channel estimation, and decoding. 
At a blocklength of $64$ complex symbols and a rate of one bit per channel use, the proposed TurboAE achieves a lower \ac{BER} than all evaluated pilot-assisted 5G baselines, with gains of 
\qty{1.9}{\decibel} at a \ac{BER} of $10^{-3}$ over a baseline involving a \ac{LDPC} code and \qty{1.7}{\decibel} over a baseline involving a polar code and \acl{SCL} decoding (list size $8$). 
In terms of \ac{BLER}, the TurboAE outperforms 
the LDPC-coded baseline, but lags behind the polar-coded baseline by about \qty{0.8}{\decibel} at a \ac{BLER} of $10^{-3}$. 
Moreover, all evaluated schemes remain at least \qty{2.1}{\decibel} away from a pilot-assisted achievability bound, indicating substantial room for improved designs.
\end{abstract}

\acresetall


\section{Introduction}
Modern coding schemes, such as \ac{LDPC} and polar codes~\cite{gallagerLowdensityParitycheckCodes1962,arikanChannelPolarizationMethod2009}, achieve excellent error-correction performance at long blocklengths. However, their performance degrades significantly in the short-blocklength regime, for two main reasons. First, short blocklengths limit the ability of these codes to fully exploit their structural properties. Second, when the blocklength is short, the finite symbol budget must be shared among synchronization, channel estimation, and data transmission. This introduces a critical trade-off between pilot symbols and payload in conventional pilot-assisted communication systems.

Finite-blocklength information-theoretic analyses show that short-packet transmission can benefit from joint synchronization, packet detection, channel estimation, and decoding. Specifically, joint synchronization and decoding without pilots achieves nearly the same coding rate as decoding with perfect synchronization, which suggests that pilot-assisted schemes may incur a rate loss~\cite{polyanskiyAsynchronousCommunication2013}. Similarly, joint packet detection and decoding achieves higher coding rates than pilot-based detection followed by decoding~\cite{lanchoJointDetectionDecoding2021}, and noncoherent transmission without explicit channel estimation is more energy efficient than pilot-assisted transmission~\cite{ostmanShortPackets2019}. However, how to design practical joint schemes that realize these gains remains an open question.

One promising approach to optimize the transmitter and receiver for short-packet communication is end-to-end learning using autoencoders \cite{osheaIntroductionDeepLearning2017}. 
However, the original autoencoder proposed in~\cite{osheaIntroductionDeepLearning2017} is limited to relatively short messages of up to about $10$ bits. 
To solve this problem, more sophisticated architectures have been proposed, such as the \ac{TurboAE}~\cite{jiangTurboAutoencoderDeep2019a} and its serial extension~\cite{clausiusSerialVsParallel2021}, inspired by the parallel/serial encoding and iterative decoding of turbo codes~\cite{berrouShannonLimitErrorcorrecting1993,669119}. In the analysis of both schemes, however, perfect frame synchronization was assumed~\cite{jiangTurboAutoencoderDeep2019a,clausiusSerialVsParallel2021}.
This assumption was later removed in~\cite{dornerLearningJointDetection2022}, in which the serial \ac{TurboAE} is extended to unsynchronized multipath channels, where it jointly performs packet detection, synchronization, equalization, and decoding. The TurboAE in \cite{dornerLearningJointDetection2022} outperforms a pilot-assisted \ac{LDPC} baseline for short blocklengths up to around $100$ channel uses. However, it remains unclear how close the performance of such a learned scheme is to information-theoretic bounds.

In this paper, we address this question by considering 
a continuous-time waveform channel with an unknown delay and memoryless block fading, for which a finite-blocklength pilot-assisted achievability bound is available~\cite{kislalSynchronizationBottleneckPilotAssisted2024b}. 
We adapt the TurboAE in \cite{dornerLearningJointDetection2022} to this channel by extending the encoder, synchronizer, decoder, and training strategy. 
The resulting learned scheme achieves a lower \ac{BER} than all evaluated pilot-assisted 5G baselines. However, both the \ac{TurboAE} and the 5G baselines remain at least \qty{2.1}{\decibel} from the finite-blocklength bound for a \ac{BLER} of $10^{-3}$, indicating substantial room for improved designs.

\section{System Model}
\label{sec:system}

\begin{figure}[t]
\centering

\begin{tikzpicture}[
  scale=1.0,
  transform shape,
  x=1cm,
  y=1cm,
  font=\footnotesize,
  text=e2eDark,
  >=latex,
  signal/.style={->,draw=e2eDark,line width=.72pt},
  module/.style={
    rounded corners=1.0pt,
    line width=.65pt,
    align=center,
    inner sep=2pt
  },
  add/.style={
    circle,
    draw=e2eDark,
    fill=white,
    line width=.55pt,
    minimum size=.40cm,
    inner sep=0pt,
    font=\footnotesize,
    execute at begin node={$+$}
  },
  multiple/.style={
    circle,
    draw=e2eDark,
    fill=white,
    line width=.55pt,
    minimum size=.40cm,
    inner sep=0pt,
    font=\footnotesize,
    execute at begin node={$\times$}
  },
  vdots/.style={
    draw=none,
    fill=none,
    minimum size=.40cm,
    inner sep=0pt,
    font=\footnotesize,
    execute at begin node={$\vdots$}
  },
  delay/.style={
    rectangle,
    rounded corners=1.0pt,
    draw=e2eDark,
    fill=white,
    line width=.55pt,
    minimum width=0.80cm,
    minimum height=0.80cm,
    inner sep=1pt,
    font=\footnotesize
  },
  pulse/.style={
    rectangle,
    rounded corners=1.0pt,
    draw=e2eDark,
    fill=white,
    line width=.55pt,
    minimum width=1.05cm,
    minimum height=0.8cm,
    inner sep=1pt,
    font=\footnotesize
  },
]


\node[font=\footnotesize] at (-1.19,1.4){$\vecu$};

\node[module,rotate=90,draw=e2ePurple,fill=e2ePurpleFill,
      minimum width=3.80cm,minimum height=0.42cm] (Encoder) at (-0.8,1.70) {\text{Encoder}};
\node[font=\footnotesize] at (-0.40,1.4){$\vecx$};

\node[module,draw=e2eGold,fill=e2eGoldFill,
      minimum width=6.50cm,minimum height=3.80cm] (channel) at (3.03,1.70) {};
\node[font=\footnotesize] at (0.19,2.8){$\vecx_1$};
\node[font=\footnotesize] at (0.19,0.8){$\vecx_{n_b}$};
\node[pulse, align=center, font=\footnotesize] (pulse1) at (1.035,2.50) {Pulse\\[0pt]Shaping};
\node[pulse, align=center, font=\footnotesize] (pulse2) at (1.035,0.50) {Pulse\\[0pt]Shaping};
\node[delay,align=center, font=\footnotesize] (delay1) at (2.90,2.50) {Delay\\[0pt]$d_1$};
\node[delay,align=center, font=\footnotesize] (delay2) at (2.90,0.50) {Delay\\[0pt]$d_{n_b}$};
\node[font=\footnotesize] at (2.03,2.8){$\rndx_1(t)$};
\node[font=\footnotesize] at (2.03,0.8){$\rndx_{n_b}(t)$};

\node[multiple] (multiple1) at (3.84,2.50) {};
\node[multiple] (multiple2) at (3.84,0.50) {};
\node[font=\footnotesize] at (3.80,3.35){$H_1$};
\node[font=\footnotesize] at (3.80,1.35){$H_{n_b}$};
\node[add] (add1) at (4.66,2.50) {};
\node[add] (add2) at (4.66,0.50) {};
\node[font=\footnotesize] at (4.68,3.35){$Z_1 (t)$};
\node[font=\footnotesize] at (4.68,1.35){$Z_{n_b} (t)$};
\node[font=\footnotesize] at (5.32,2.8){$\rndy_1(t)$};
\node[font=\footnotesize] at (5.32,0.8){$\rndy_{n_b}(t)$};

\node[vdots] (vdots) at (1.035,1.70) {};
\node[vdots] (vdots) at (2.90,1.70) {};
\node[vdots] (vdots) at (4.25,1.93) {};
\node[module,rotate=90,draw=e2eDark,fill=white,line width=.55pt,font=\footnotesize, minimum width=3.20cm,minimum height=0.42cm] (filtering) at (5.99,1.60) {\text{Filtering \& Oversampling}};
\node at (6.46,1.40){$\matY$};

\node[module,rotate=90,draw=e2eBlue!70,fill=e2eBlueFill,
      minimum width=3.80cm,minimum height=0.42cm] (Decoder) at (6.85,1.70) {\text{Decoder}};
\node[font=\footnotesize] at (7.24,1.4){$\hat{\vecu}$};

\draw[signal] ([xshift=-10pt]Encoder.north) --(Encoder.north);
\draw[signal] (Encoder.south) --([xshift=3pt]channel.west);
\draw[signal] (Encoder.south) -- (-0.13,1.70) -- (-0.13,2.50) -- (pulse1.west);
\draw[signal] (Encoder.south) -- (-0.13,1.70) -- (-0.13,0.50) -- (pulse2.west);

\draw[signal] (pulse1.east) --(delay1.west);
\draw[signal] (pulse2.east) --(delay2.west);

\draw[signal] (delay1.east) --(multiple1.west);
\draw[signal] (delay2.east) --(multiple2.west);
\draw[signal] ([yshift=12pt]multiple1.north) -- (multiple1.north);
\draw[signal] ([yshift=12pt]multiple2.north) -- (multiple2.north);

\draw[signal] (multiple1.east) --(add1.west);
\draw[signal] (multiple2.east) --(add2.west);
\draw[signal] ([yshift=12pt]add1.north) -- (add1.north);
\draw[signal] ([yshift=12pt]add2.north) -- (add2.north);

\draw[signal] (add1.east) --(add1.east -| filtering.north);
\draw[signal] (add2.east) --(add2.east -| filtering.north);
\draw[signal] (filtering.south |- Decoder.north) --(Decoder.north);
\draw[signal] (Decoder.south) --([xshift=10pt]Decoder.south);

\end{tikzpicture}
\vspace{-15pt}
\caption{Block diagram of the system model. Each symbol is pulse shaped. The resulting waveform is delayed by $d_\ell$, $\ell=1,\ldots,n_b$, scaled by the fading coefficient $H_\ell$, and corrupted by noise before being filtered and oversampled at the receiver.}
\label{fig:system_model_2}
\end{figure}
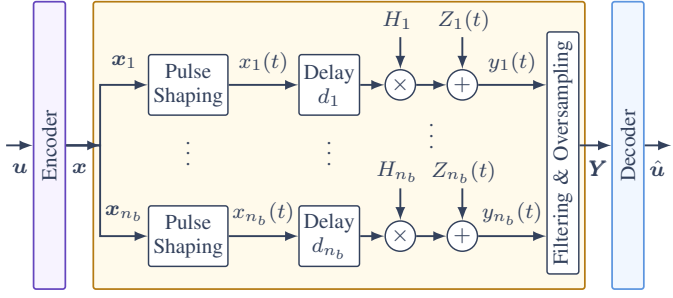
We use the same single-input single-output waveform model as in~\cite{kislalSynchronizationBottleneckPilotAssisted2024b}, which is illustrated in Fig.~\ref{fig:system_model_2}. The goal is to transmit~$k$ information bits $\vecu \in \{0,1\}^k$ within a packet $\vecx \in \mathbb{C}^{n}$ comprising $n$ complex-valued channel uses. The packet is divided into $n_b$ blocks, each containing $n_c$ symbols, such that $n = n_b n_c$. Each block is transmitted over an independent fading channel, with fading coefficient remaining constant over the block. The vector of transmitted symbols corresponding to the $\ell$th block is
\begin{equation}
\vecx_\ell =
\left[x_{1,\ell},\ldots,x_{n_c,\ell}\right]^\transpose
\in \mathbb{C}^{n_c}, \quad \ell=1,\ldots,n_b .
\end{equation}
The full transmitted symbol vector is
$\vecx=[\vecx_1^\transpose, \ldots, \vecx_{n_b}^\transpose]^\transpose$. We assume that the energy of each transmitted complex symbol is normalized to $1$, i.e.,
\begin{equation}
\left|x_{i,\ell}\right|^2=1, \quad i=1,\ldots,n_c,\quad \ell=1,\ldots,n_b.
\label{eq:unit_power_constraint}
\end{equation}
The discrete symbols are converted into a continuous-time waveform via pulse shaping as follows:
\begin{equation}
\rndx_\ell(t)=\sum_{i=1}^{n_c} x_{i,\ell} s_T(t-(i-1)T).
\label{eq:continuous_waveform}
\end{equation}
Here, $T$ is the symbol duration, and $s_T(t)$ is a unit-energy rectangular pulse
\begin{equation}
s_T(t)=
\begin{cases}
1/\sqrt{T}, & 0 \leq t < T,\\
0, & \text{otherwise}.
\end{cases}
\label{eq:pulse_equation}
\end{equation}
The transmitted waveform $\rndx_\ell(t)$ is then passed through a fading channel with unknown delay and fading gain. The corresponding waveform is
\begin{equation}
\rndy_\ell(t)=H_\ell \rndx_\ell(t-d_\ell) + Z_\ell(t),
\label{eq:received_waveform_signal}
\end{equation}
where $H_\ell \stackrel{\mathrm{i.i.d.}}{\sim}\mathcal{CN}(0,1)$ denotes the fading coefficient for block \(\ell\) and $d_\ell \sim \mathcal{U}(0,d_{\max})$ is the unknown uniformly distributed delay. Throughout, we consider the common delay case $d_1=\cdots=d_{n_b}=d$. Finally, $Z_\ell(t)$ in~\eqref{eq:received_waveform_signal} is a complex white Gaussian noise process with power spectral density $N_0$. Under the unit-symbol-energy normalization~\eqref{eq:unit_power_constraint}, the average \ac{SNR} per complex symbol is $\rho = 1/N_0$.

Following \cite{kislalSynchronizationBottleneckPilotAssisted2024b}, we assume that, to obtain an oversampled observation, the receiver filters $\rndy_\ell(t)$ with the unit-energy rectangular pulse $s_{t_s}(t)$, where $t_s=T/N$ and $N$ is the oversampling rate, and samples the filter output every $t_s$ seconds. The $m$th received sample in block~$\ell$ is
\begin{equation}
y_{m,\ell}=(\rndy_\ell*s_{t_s})(m t_s),\qquad m=1,\ldots,M,
\end{equation}
where $M=\left\lceil d_{\max}/t_s\right\rceil+n_cN$. Note that $M$ is larger than $n_cN$ to account for the unknown delay. We collect these samples in the vector
\begin{equation}
\vecy_\ell=[y_{1,\ell},\ldots,y_{M,\ell}]^\transpose\in\mathbb{C}^{M}.
\label{eq:oversampled_received_vector}
\end{equation}

It turns out convenient to write $d$ as $(q+\epsilon)t_s$, where $q=\lfloor d/t_s\rfloor$ is the integer delay in samples, $e=d-q t_s$ is the fractional delay, and $\epsilon=e/t_s\in[0,1)$ is the normalized fractional delay. To capture oversampling, we next define
\begin{equation}
x_{\mathlabel{N},r,\ell}=x_{\lceil r/N\rceil,\ell}, \qquad r=1,\ldots,Nn_c,
\label{eq:oversampling}
\end{equation}
and use~\eqref{eq:oversampling} to construct the following scaled, zero-padded vector shifted by $q$ samples:
\begin{equation}
\vecx_{\mathlabel{N},\ell}(q)=\frac{1}{\sqrt{N}}
\left[
\veczero_q^\transpose,
x_{\mathlabel{N},1,\ell},\ldots,
x_{\mathlabel{N},Nn_c,\ell},
\veczero_{M-q-Nn_c}^\transpose
\right]^\transpose.
\label{eq:shifted_symbol_vector}
\end{equation}
Using~\eqref{eq:shifted_symbol_vector}, we can express $\vecy_\ell$  in~\eqref{eq:oversampled_received_vector} as~\cite[Sec.~II-C]{kislalSynchronizationBottleneckPilotAssisted2024b}
\begin{equation}
\vecy_\ell
=
H_\ell\left[
(1-\epsilon)\vecx_{\mathlabel{N},\ell}(q)
+\epsilon\vecx_{\mathlabel{N},\ell}(q+1)
\right]
+\rvecz_\ell,
\label{eq:received_sample_model}
\end{equation}
where $\rvecz_\ell\sim\mathcal{CN}(\veczero_M,\rho^{-1}\matI_M)$. The noise samples are independent because the unit-energy receive filters span nonoverlapping sampling intervals. Stacking the $\{\vecy_\ell\}$, we obtain the received signal (see Fig.~\ref{fig:system_model_2})
\begin{equation}
\matY=[\vecy_1,\ldots,\vecy_{n_b}]^\transpose\in\mathbb{C}^{n_b\times M}.
\end{equation}

\begin{figure*}[t]
\centering
\input{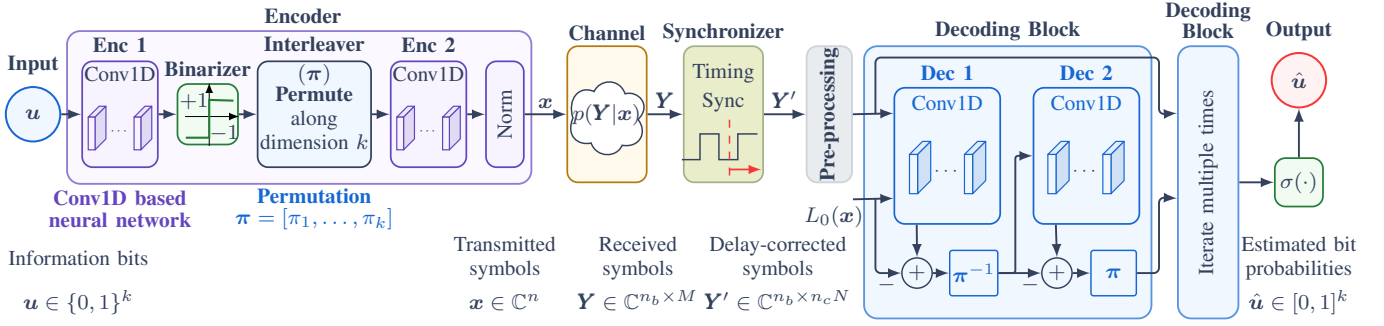}
\vspace{-15pt}
\caption{End-to-end TurboAE architecture.}
\label{fig:end2end_turboae}
\end{figure*}

\section{Finite-blocklength Bound \& 5G Baselines}
\label{sec:baseline}
%
%
%
Both the finite-blocklength bound and the 5G baselines are obtained under the assumption that each fading block carries a common pilot vector $\vecx^{(p)}$ of dimension $n_p$, followed by a data vector~$\vecx^{(d)}_\ell$ of dimension $n_s=n_c-n_p$, i.e.,
\begin{equation}
\vecx_\ell=[(\vecx^{(p)})^\transpose,(\vecx^{(d)}_\ell)^\transpose]^\transpose.
\label{eq:pilot_assisted_transmitted_vector}
\end{equation}
The receiver estimates the delay and fading gains from the pilot vectors. Then, it decodes the data by treating these estimates as perfect.

\subsection{Synchronization and Channel Estimation}
\label{sec:ml_sync}
Since we assume that the pilots precede the data in~\eqref{eq:pilot_assisted_transmitted_vector}, all pilot samples of block~$\ell$ lie within the first $M_p=\lceil d_{\max}/t_s\rceil+n_pN$ entries of $\vecy_\ell$ in~\eqref{eq:oversampled_received_vector}. We denote these samples by $\vecy^{(p)}_\ell\in\mathbb{C}^{M_p}$. As in~\cite[Sec.~II-C]{kislalSynchronizationBottleneckPilotAssisted2024b}, we consider an estimator that neglects the interference between data symbols and pilots. For a candidate integer delay $\bar q$ and fractional delay $\bar\epsilon$, we define the noiseless pilot vector per unit channel gain as
\begin{equation}
\vecv(\bar q,\bar\epsilon)=(1-\bar\epsilon)\,\vecx^{(p)}_{\mathlabel{N}}(\bar q)+\bar\epsilon\,\vecx^{(p)}_{\mathlabel{N}}(\bar q+1),
\label{eq:noiseless_pilot_vector}
\end{equation}
where $\vecx^{(p)}_{\mathlabel{N}}(\bar q)$ is defined similarly as $\vecx_{\mathlabel{N},\ell}(q)$ in~\eqref{eq:shifted_symbol_vector}, with $n_c$ and $M$ replaced by~$n_p$ and~$M_p$, respectively. Using~\eqref{eq:received_sample_model} and~\eqref{eq:noiseless_pilot_vector}, we conclude that $\vecy^{(p)}_\ell=H_\ell\vecv(q,\epsilon)+\rvecz^{(p)}_\ell$, where $\rvecz^{(p)}_\ell\sim\mathcal{CN}(\veczero,\rho^{-1}\matI_{M_p})$. We use a \ac{ML} estimator to jointly estimate the fading gains, integer delay, and fractional delay as
\begin{equation}
[\hat\vech,\hat q,\hat\epsilon]=\argmin_{\bar\vech,\bar q,\bar\epsilon}\,\sum_{\ell=1}^{n_b}\bigl\|\vecy^{(p)}_\ell-\bar h_\ell\,\vecv(\bar q,\bar\epsilon)\bigr\|^2.
\label{eq:joint_ml}
\end{equation}
For fixed $(\bar q,\bar\epsilon)$, the conditional gain estimate for block~$\ell$ is
\begin{equation}
\hat H_\ell(\bar q,\bar\epsilon)=\frac{\vecv(\bar q,\bar\epsilon)^H\vecy^{(p)}_\ell}{\|\vecv(\bar q,\bar\epsilon)\|^2}.
\label{eq:conditional_gain_estimator}
\end{equation}
Substituting~\eqref{eq:conditional_gain_estimator} into~\eqref{eq:joint_ml}, we obtain the common-delay estimate
\begin{equation}
[\hat q,\hat\epsilon]=\argmax_{\bar q,\bar\epsilon}\,\sum_{\ell=1}^{n_b}\frac{\bigl|\vecv(\bar q,\bar\epsilon)^H\vecy^{(p)}_\ell\bigr|^2}{\|\vecv(\bar q,\bar\epsilon)\|^2}.
\label{eq:concentrated_ml}
\end{equation}
For fixed $\bar q$, the objective in~\eqref{eq:concentrated_ml} is a rational function of~$\bar\epsilon$ with quadratic numerator and denominator. Hence, the optimal~$\bar\epsilon$ is $0$, $1$, or a point in $(0,1)$ obtained by setting the derivative of~\eqref{eq:concentrated_ml} to zero. Evaluating these candidates for all $\bar q$, we obtain $(\hat q,\hat\epsilon)$~\cite[Alg.~1]{kislalSynchronizationBottleneckPilotAssisted2024b}. Finally, $\hat d=(\hat q+\hat\epsilon)t_s$ and $\hat H_\ell=\hat H_\ell(\hat q,\hat\epsilon)$.
\subsection{Finite-blocklength Achievability Bound}
\label{sec:rcus}
We consider a receiver that filters $\rndy_\ell(t)$ in~\eqref{eq:received_waveform_signal} with $s_T(t)$ and resamples at the symbol rate using the delay estimate $\hat d$, obtaining the data samples $y^{(d)}_{i,\ell}=(\rndy_\ell*s_T)((n_p+i)T+\hat d)$, $i=1,\ldots,n_s$. Then, the transmitted codeword is recovered by a mismatched \ac{SNN} decoder, which treats the estimated fading coefficients as correct and returns
\begin{equation}
\hat\vecx^{(d)}=\argmin_{\bar\vecx\in\setC}\,\sum_{\ell=1}^{n_b}\bigl\|\vecy^{(d)}_\ell-\hat H_\ell\,\bar\vecx_\ell\bigr\|^2,
\end{equation}
where $\setC$ contains the $2^k$ codewords. We use the so-called \ac{RCUs} from finite-blocklength information theory.
Specifically, we assume that the codeword symbols are generated as $x_{i,\ell}=e^{\mathrm{j}\theta_{i,\ell}}$, where $\theta_{i,\ell}\stackrel{\mathrm{i.i.d.}}{\sim}\mathcal{U}(0,2\pi)$. Define the event that transmitted symbols are synchronized to within one symbol period as $\mathcal{A}=\{|\hat d-d|\leq T\}$, and let $\mathcal{A}^c$ denote its complement. The bound is stated in the following theorem~\cite[Sec.~III]{kislalSynchronizationBottleneckPilotAssisted2024b}. 
\begin{theorem}[Pilot-assisted \ac{RCUs} bound]
\label{thm:rcus}
For the pilot-assisted model above and every $s>0$, there exists a codebook with $2^k$ codewords, each containing $n_bn_s$ data symbols, whose average packet error probability $P_{\mathlabel{e}}$, averaged over equiprobable messages, satisfies
\begin{equation}
P_{\mathlabel{e}}\leq\opP[\mathcal{A}]\epsilon_{\mathlabel{ub}}(s)+\opP[\mathcal{A}^{c}],
\label{eq:rcus}
\end{equation}
where
\begin{equation}
\epsilon_{\mathlabel{ub}}(s)=\opE\!\left[\min\left\{1,e^{k\ln2-I_s}\right\}\,\middle|\,\mathcal{A}\right].
\label{eq:conditional_rcus}
\end{equation}
Here, $s$ is an optimization parameter, and 
\begin{equation}
I_s=\sum_{\ell=1}^{n_b}\sum_{i=1}^{n_s}\log\frac{e^{-s|y^{(d)}_{i,\ell}-\hat H_\ell x^{(d)}_{i,\ell}|^2}}{\opE_{\bar x}\bigl[e^{-s|y^{(d)}_{i,\ell}-\hat H_\ell\bar x|^2}\bigr]},
\label{eq:generalized_information_density}
\end{equation}
where $\bar x$ follows the same distribution as $x_{i,\ell}$ and is independent of all other quantities.
\end{theorem}


\subsection{Pilot-Assisted 5G Baselines}
\label{sec:5G_baseline}
Similar to the \ac{RCUs} bound, the baseline receiver uses the delay and fading gain estimates from Section~\ref{sec:ml_sync}, filters the waveform with $s_T(t)$, resamples at the symbol rate, and decodes by treating the estimated delay and channel coefficients as perfect. We compare three coding schemes: a 5G polar code with \ac{CRC}-aided  successive-cancellation (SC) decoding (5G polar + CRC SC), a 5G polar code with \ac{CRC}-aided \ac{SCL} decoding of list size eight (5G polar + CRC SCL-8), and a 5G \ac{LDPC} code with 20 belief propagation iterations (5G LDPC BP-20). To satisfy the constraint~\eqref{eq:unit_power_constraint}, we assume that all coding schemes use \ac{PSK} modulation. 

\section{TurboAE Design}
\label{sec:turboae}
The proposed \ac{TurboAE}, illustrated in Fig.~\ref{fig:end2end_turboae}, consists of a learned encoder at the transmitter and a learned synchronizer and decoder at the receiver. While the encoder and decoder largely build on~\cite[Figs.~5 and~20]{dornerLearningJointDetection2022}, our main contributions, i.e., the synchronizer and the training strategy, differ from~\cite{dornerLearningJointDetection2022} in two respects. 
First, the synchronizer in~\cite{dornerLearningJointDetection2022} represents the integer delay as a one-hot probability vector and lets the fractional delay to be accounted for by the decoder. 
For the assumed rectangular pulse in~\eqref{eq:pulse_equation}, this design would lead to significantly degraded performance. 
Our synchronizer instead estimates both integer and fractional delay and compensates for them before decoding (see Section~\ref{sec:sync}). 
Second, in the training strategy used in~\cite{dornerLearningJointDetection2022}, the synchronizer is trained jointly with the decoder using a categorical cross-entropy loss. 
To suit the proposed synchronizer design, we instead use a regression loss for synchronization and perform the training in two stages: first we train the encoder, decoder, and synchronizer separately; then jointly (see Section~\ref{sec:training}). We describe each component of the proposed architecture next.

\subsection{Encoder}
\label{sec:encoder}
The encoder contains two 1D convolutional blocks, separated by a straight-through binarizer~\cite{bengioEstimatingPropagatingGradients2013} and a fixed interleaver. The outputs of the encoder are normalized to satisfy~\eqref{eq:unit_power_constraint}.

\subsection{Synchronizer}
\label{sec:sync}

\begin{figure}[t]
\centering
\usetikzlibrary{calc}
\begin{tikzpicture}[
  x=1cm,y=1cm,font=\footnotesize,text=black!85,>=latex,
  signal/.style={->,draw=black!65,line width=.65pt},
  wire/.style={draw=black!65,line width=.65pt},
  module/.style={
    rectangle,rounded corners=2.5pt,line width=.65pt,
    align=center,inner xsep=3pt,inner ysep=3pt,anchor=north
  },
  learned/.style={module,draw=e2eBlue,fill=e2eBlue!6},
  fixed/.style={module,draw=black!40,fill=black!4},
  edgelabel/.style={fill=white,inner xsep=2pt,inner ysep=1pt},
  junction/.style={circle,fill=black!65,inner sep=0pt,minimum size=2.5pt}
]

\node[anchor=west,inner sep=2pt] (input) at (-4.03,0)
  {\textbf{Input:} $\matY\in\mathbb{C}^{n_b\times M}$};
\node[learned,minimum width=.26cm,minimum height=.20cm,inner sep=0pt,anchor=center]
  at (.05,0) {};
\node[anchor=west,inner sep=2pt] at (.22,0) {trainable};
\node[fixed,minimum width=.26cm,minimum height=.20cm,inner sep=0pt,anchor=center]
  at (1.85,0) {};
\node[anchor=west,inner sep=2pt] at (2.02,0) {closed form};

\coordinate (center) at (0,0);
\coordinate (split) at (input.south|-0,-.32);
\coordinate (rcol) at (2.20,0);
\draw[wire] (input.south)--(split);
\node[junction] at (split) {};

\node[fixed,text width=3.75cm,minimum height=1.15cm] (stats) at (-2.20,-.66)
  {\textbf{Candidate features}~\eqref{eq:sync_prior}\\[2pt]
   one feature vector $\veca_q$ per\\ integer-delay hypothesis $q$};

\node[learned,text width=3.75cm,minimum height=1.15cm] (heads) at ([yshift=-.45cm]stats.south)
  {\textbf{Embedding \& heads}~\eqref{eq:sync_heads}\\[2pt]
   $\veca_q\rightarrow\vecz_q$, then\\ integer and fractional heads};

\draw[signal] (split)--(stats.north|-split)--(stats.north);
\draw[signal] (stats.south)--(heads.north)
  node[midway,right] {$\veca_q$};

\coordinate (lcenter) at ($(stats.north)!0.5!(heads.south)$);
\node[learned,text width=3.75cm,anchor=center,minimum height=1.5cm] (temporal) at (lcenter-|rcol)
  {\textbf{Global temporal features}~\eqref{eq:sync_fractional_estimate}\\[2pt]
   learned features of $\matY$, pooled\\ over blocks and samples into $\vecg$\\[2pt]
   $[\vece^{(0)};\vecg]\rightarrow\Delta e^{\mathlabel{glob}}$};

\draw[signal] (split)--(temporal.north|-split)--(temporal.north);

\draw[wire] (stats.east)--(stats.east-|center);
\draw[signal] (stats.east-|center)--(temporal.west-|center)--(temporal.west);
\path (stats.east-|center)--(temporal.west-|center)
  node[pos=.5,edgelabel,rotate=90] {$\vece^{(0)}$};

\node[fixed,text width=8.15cm,minimum height=.60cm] (fusion) at ([yshift=-.47cm]heads.south-|center)
  {\textbf{Soft combination}~\eqref{eq:soft_delay_estimate}\quad
   \mbox{$\hat d=\sum_{q\in\mathcal{Q}} p_q(\matY)(qt_s+\hat e_q)$}};
\draw[signal] (heads.south)--(heads.south|-fusion.north)
  node[midway,right] {$p_q(\matY),\,e_q^{(1)}$};
\draw[signal] (temporal.south)--(temporal.south|-fusion.north)
  node[pos=.62,right] {$\Delta e^{\mathlabel{glob}}$};

\node[fixed,text width=8.15cm,minimum height=.60cm] (correct) at ([yshift=-.42cm]fusion.south)
  {\textbf{Delay correction}~\eqref{eq:sync_pulse_matrix}\quad
   \mbox{discard $\hat q$ samples, regularized least squares}};
\draw[signal] (fusion.south)--(correct.north) node[midway,right] {$\hat d$};

\node[anchor=north,inner sep=2pt] (output) at ([yshift=-.30cm]correct.south)
  {\textbf{Output:} $\matY'\in\mathbb{C}^{n_b\times n_cN}$};
\draw[signal] (correct.south)--(output.north);
\end{tikzpicture}
\vspace{-5pt}
\caption{Learned model-guided neural synchronizer. Candidate features provide integer-delay probabilities and fractional-delay priors, while global temporal features refine the fractional-delay estimates before soft combination.}
\label{fig:sync_block}
\end{figure}
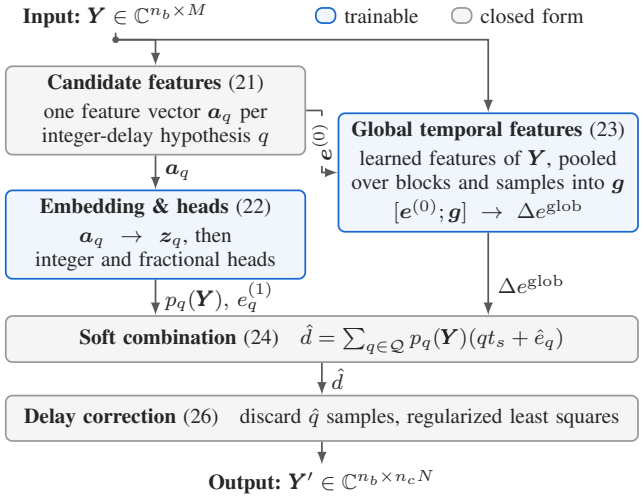

The learned synchronizer, illustrated in Fig.~\ref{fig:sync_block}, maps the oversampled observations $\matY \in \mathbb{C}^{n_b\times M}$ to one continuous common-delay estimate $\hat d$ and uses it to obtain the delay-corrected observations ${\matY^\prime}\in\mathbb{C}^{n_b\times n_cN}$. It builds on two key ideas. First, instead of predicting $d$ directly, it predicts the integer and fractional delays ($q$ and $e$) separately and combines them into $\hat{d}$. Second, rather than predicting $e$ in a single step, it first forms a model-based prior estimate by conventional feature extraction and then corrects it with neural networks. We next describe each block in Fig.~\ref{fig:sync_block}.

\paragraph{Candidate features}
\label{para:candidate_features}
For each possible integer delay $q\in\mathcal{Q}=\{0,\ldots,\lfloor d_{\max}/t_s\rfloor\}$, let $\mathcal{W}_q=\mbox{$\{q+1,\ldots,q+n_cN+1\}$}$ index the samples the packet would occupy if the integer delay were $q$. We consider the following nine candidate features $\veca_q\in\mathbb{R}^{9}$: the normalized assumed delay $q/\lfloor d_{\max}/t_s\rfloor$; three \emph{power features}, i.e., the empirical mean and standard deviation of the power of the samples inside $\mathcal{W}_q$ and the alignment score $s_q$, defined as the average power inside $\mathcal{W}_q$ minus the power outside $\mathcal{W}_q$; two \emph{edge features}, i.e., ${E_{\mathlabel{L},q}+E_{\mathlabel{R},q}}$ and $e_q^{(0)}/t_s$, where $E_{\mathlabel{L},q}/E_{\mathlabel{R},q}$ is the difference between the power of the first/last sample inside $\mathcal{W}_q$ and the average power outside $\mathcal{W}_q$; and
\begin{equation}
    e_q^{(0)}=\frac{t_s\sqrt{E_{\mathlabel{R},q}}}
    {\sqrt{E_{\mathlabel{L},q}}+\sqrt{E_{\mathlabel{R},q}}}
\label{eq:sync_prior}
\end{equation}
is an estimate of the fractional-delay prior (see~\eqref{eq:received_sample_model}); 
and three \emph{correlation features}, i.e., the real part, imaginary part, and magnitude of the mean of $y_{m+1,\ell}\,y_{m,\ell}^*$ inside $\mathcal{W}_q$.

\paragraph{Embedding \& heads}
We use a two-layer fully connected network to embed $\veca_q \in \mathbb{R}^{9}$ into $\vecz_q \in \mathbb{R}^{64}$. Furthermore, we use two separate linear heads ($f_{\mathlabel{I}}$ and $f_{\mathlabel{E}}$) combined with $\tanh$ activation to produce the integer-delay probabilities $p_q(\matY)$ and initial fractional delay $e_q^{(1)}$, 
where $\gamma$ is a trainable scalar and the $\tanh$ terms determine the influence of the correction according to
\begin{equation}
\begin{aligned}
    p_q(\matY)&=\operatorname*{softmax}_{q\in\mathcal{Q}}
    \bigl(f_{\mathlabel{I}}(\vecz_q)+\gamma\tanh(s_q)\bigr),\\
    e_q^{(1)}&=e_q^{(0)}+t_s\tanh\bigl(f_{\mathlabel{E}}(\vecz_q)\bigr).
\end{aligned}
\label{eq:sync_heads}
\end{equation}

\paragraph{Global temporal features}
In parallel, we use another neural network $f_{\mathlabel{G}}$, consisting of residual 1D convolutional blocks and an attention-pooling layer, to learn temporal features $\vecg$ directly from $\matY$ as complementary information for the fractional-delay correction $\Delta e^{\mathlabel{glob}}$ according to
\begin{equation}
    \Delta e^{\mathlabel{glob}}=t_s\tanh\bigl(f_{\mathlabel{G}}([\vece^{(0)};\vecg])\bigr),
    \quad \vece^{(0)}=[e_q^{(0)}]_{q\in\mathcal{Q}}.
\label{eq:sync_fractional_estimate}
\end{equation}

\paragraph{Soft combination}
Combining~\eqref{eq:sync_heads} with the refined fractional-delay estimates $\hat e_q=e_q^{(1)}+\Delta e^{\mathlabel{glob}}$, we obtain the final common delay estimate
\begin{equation}
    \hat d=\sum_{q\in\mathcal{Q}} p_q(\matY)(qt_s+\hat e_q).
\label{eq:soft_delay_estimate}
\end{equation}

\paragraph{Delay correction}
Let $\hat q=\lfloor\hat d/t_s\rfloor$ and $\hat\epsilon=\hat d/t_s-\hat q$. In block $\ell$, we remove the first $\hat q$ samples that carry only noise and retain $\widetilde{\vecy}_\ell=[y_{\hat q+1,\ell},\ldots,y_{\hat q+n_cN+1,\ell}]^\transpose$. Then, we undo the interference between neighboring symbols caused by the fractional delay. By~\eqref{eq:received_sample_model}, this interference is described by
\begin{equation}
    \matB=(1-\hat\epsilon)\begin{bmatrix}\matU\\\veczero^\transpose\end{bmatrix}
    +\hat\epsilon\begin{bmatrix}\veczero^\transpose\\\matU\end{bmatrix},
    \qquad \matU=\matI_{n_c}\otimes\vecone_N,
\end{equation}
where $\matU\in\mathbb{R}^{n_cN\times n_c}$ repeats each symbol $N$ times, $\matI_{n_c}$ is the identity matrix, $\vecone_N$ is a column vector of $N$ ones, and~$\otimes$ denotes the Kronecker product. The zero row $\veczero^\transpose$ has $n_c$ entries. Treating $\hat d$ as exact, we recover the faded symbols by regularized least squares with regularization parameter $\nu=0.2$ and repeat them according to the oversampling factor $N$:
\begin{equation}
    \vecy'_\ell=\matU(\matB^\transpose\matB+\nu\matI_{n_c})^{-1}
    \matB^\transpose\widetilde{\vecy}_\ell.
\label{eq:sync_pulse_matrix}
\end{equation}

\subsection{Decoder}
The decoder uses two groups of 1D convolutional networks, denoted by Dec~1 and Dec~2 in Fig.~\ref{fig:end2end_turboae}, which iteratively exchange information through the fixed interleaver $\pi$. The input $\matY'$ is first normalized and embedded along the symbol dimension, and then fed into the decoder for multiple iterations, each with its own trainable Dec~1 and Dec~2. Starting from a zero prior $L_0(\vecx)$, in each iteration, Dec~1 combines the received features with $L_i(\vecx)$, Dec~2 refines the deinterleaved output of Dec~1, and the interleaved output of Dec~2 becomes the next $L_{i+1}(\vecx)$. Each network subtracts its input information from its output, so only new (extrinsic) information is passed on.  In the last iteration, the decoder returns the $k$ bit probabilities.

\subsection{Training}
\label{sec:training}
%
%
%

Let $\boldsymbol{\theta}_{\mathlabel{Enc}}$, $\boldsymbol{\theta}_{\mathlabel{Dec}}$, and $\boldsymbol{\theta}_{\mathlabel{Sync}}$ denote the encoder, decoder, and synchronizer parameters. To train the encoder and decoder pair, we use the mean \ac{BCE} loss
\begin{equation}
    \mathcal{L}_{\mathlabel{b}}
    =-\frac{1}{k}
      \sum_{i=1}^{k}
      \left[u_{i}\log \hat u_{i}
      +(1-u_{i})\log(1-\hat u_{i})\right],
    \label{eq:turboae_bit_loss}
\end{equation}
where $u_{i}$ and $\hat u_{i}$ are the transmitted bits and corresponding predicted probabilities. To train the encoder and synchronizer pair, we use the mean smooth-$L_1$ loss
\begin{equation}
    \mathcal{L}_{\mathlabel{s}}
    =\frac{1}{n_b}\sum_{\ell=1}^{n_b}
      \psi(\hat d_{\ell}-d_{\ell}),
      \label{eq:turboae_sync_loss}
\end{equation}
where $\psi(z)=z^2/(2\beta)$ if $|z|<\beta$ and $\psi(z)=|z|-\beta/2$ otherwise, with threshold $\beta$. For joint training, we use $\mathcal{L}_{\mathlabel{J}}=\mathcal{L}_{\mathlabel{b}}+\alpha\mathcal{L}_{\mathlabel{s}}$, where the parameter $\alpha$ balances the two losses.

\begin{algorithm}[!t]
    \caption{Two-stage training of the TurboAE}
    \label{alg:turboae_training}
    \footnotesize
    \begin{algorithmic}[1]
        \Require Total steps $N_{\mathlabel{step}}$, first-stage steps
        $N_{\mathlabel{pre}}$, and pass counts $K_{\mathlabel{Enc}}$,
        $K_{\mathlabel{Dec}}$, $K_{\mathlabel{SyncEnc}}$,
        $K_{\mathlabel{Sync}}$, $K_{\mathlabel{JointEnc}}$,
        $K_{\mathlabel{JointDec}}$
        \Ensure $\boldsymbol{\theta}_{\mathlabel{Enc}}$,
        $\boldsymbol{\theta}_{\mathlabel{Dec}}$, and
        $\boldsymbol{\theta}_{\mathlabel{Sync}}$
        \State Initialize $\boldsymbol{\theta}_{\mathlabel{Enc}}$,
        $\boldsymbol{\theta}_{\mathlabel{Dec}}$,
        $\boldsymbol{\theta}_{\mathlabel{Sync}}$, and their Adam optimizers
        \For{$\mathit{i}=1,\ldots,N_{\mathlabel{step}}$}
            \If{$\mathit{i}\leq N_{\mathlabel{pre}}$}
                \Comment{stage 1: separate training}
                \State \Call{Pass}{\textsc{Enc},
                    $K_{\mathlabel{Enc}}$, $\mathcal{L}_{\mathlabel{b}}$,
                    $\{\boldsymbol{\theta}_{\mathlabel{Enc}}\}$}
                    \Comment{perfect sync.}
                \State \Call{Pass}{\textsc{Dec},
                    $K_{\mathlabel{Dec}}$, $\mathcal{L}_{\mathlabel{b}}$,
                    $\{\boldsymbol{\theta}_{\mathlabel{Dec}}\}$}
                    \Comment{perfect sync.}
                \State \Call{Pass}{\textsc{Enc},
                    $K_{\mathlabel{SyncEnc}}$, $\mathcal{L}_{\mathlabel{s}}$,
                    $\{\boldsymbol{\theta}_{\mathlabel{Enc}}\}$}
                \State \Call{Pass}{\textsc{Sync},
                    $K_{\mathlabel{Sync}}$, $\mathcal{L}_{\mathlabel{s}}$,
                    $\{\boldsymbol{\theta}_{\mathlabel{Sync}}\}$}
            \Else
                \Comment{stage 2: joint training}
                \State \Call{Pass}{\textsc{Enc+Sync},
                    $K_{\mathlabel{JointEnc}}$, $\mathcal{L}_{\mathlabel{J}}$,
                    $\{\boldsymbol{\theta}_{\mathlabel{Enc}},
                    \boldsymbol{\theta}_{\mathlabel{Sync}}\}$}
                \State \Call{Pass}{\textsc{Dec+Sync},
                    $K_{\mathlabel{JointDec}}$, $\mathcal{L}_{\mathlabel{J}}$,
                    $\{\boldsymbol{\theta}_{\mathlabel{Dec}},
                    \boldsymbol{\theta}_{\mathlabel{Sync}}\}$}
            \EndIf
        \EndFor
    \end{algorithmic}
\end{algorithm}

In Algorithm~\ref{alg:turboae_training}, we extend the alternating training of~\cite{jiangTurboAutoencoderDeep2019a}, in which the encoder is trained with the decoder frozen and vice versa to avoid poor local optima. In each \textsc{Pass}$(\cdot,K,\mathcal{L},\boldsymbol{\theta})$, we perform $K$ gradient-based updates of $\boldsymbol{\theta}$ to minimize $\mathcal{L}$, each time on a new batch of random messages and channel realizations, with all other parameters frozen. In stage one, the encoder and decoder are first trained with perfect timing to minimize the BCE loss~\eqref{eq:turboae_bit_loss}; then the encoder and synchronizer are trained to minimize the synchronization loss~\eqref{eq:turboae_sync_loss}, so that the encoder learns the two objectives alternately. In stage two, the synchronizer is updated with either the encoder or the decoder under $\mathcal{L}_{\mathlabel{J}}$, with the remaining component frozen.

\section{Results}
\label{sec:results}
\begin{figure*}[t]
\centering
\input{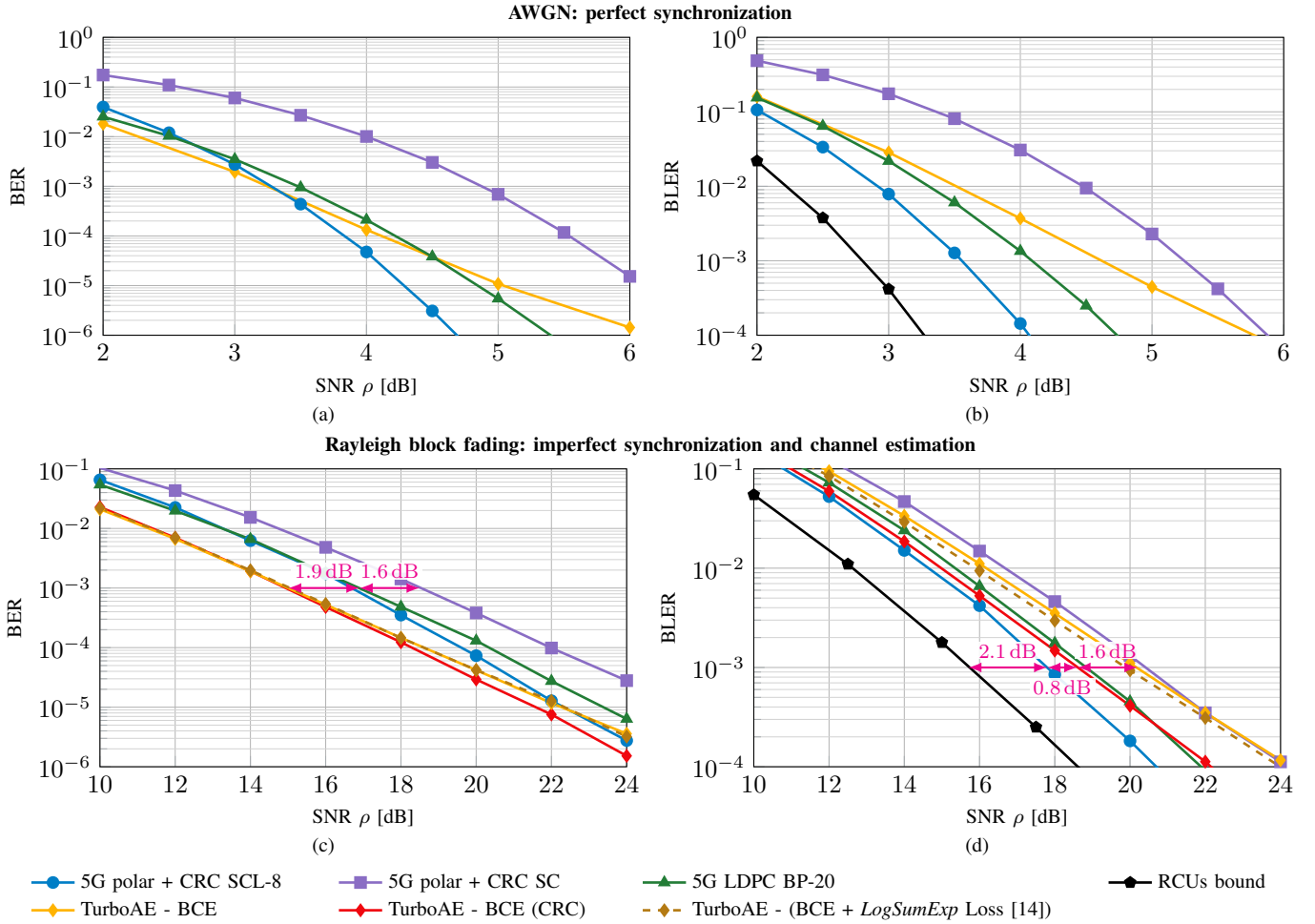}
\vspace{-8pt}
\caption{BER (left) and BLER (right) versus SNR for $n=64$ and $R=k/n=1$. Top: AWGN; bottom: Rayleigh block fading ($n_b=4$, $n_c=16$, $N=8$, common delay $d\sim\mathcal{U}(0,4T)$). The 5G baselines use QPSK under AWGN; under block fading, 5G polar + CRC SC uses 16-PSK, while 5G polar + CRC SCL-8 and 5G LDPC BP-20 use 32-PSK. Two pilots per block are transmitted in these schemes. The TurboAE uses learned symbols without dedicated pilots.}
\label{fig:ber_bler_vs_snr}
\end{figure*}

For the numerical results,\footnote{Source code, trained models, and parameter settings are available at \url{https://github.com/chenbin234/Turbo-Autoencoder-vs.-Finite-Blocklength-Bound}.} we consider $n=64$ complex symbols and a rate of $R=k/n=1$ bit per channel use, as in \cite{dornerLearningJointDetection2022}. 
We normalize the symbol period to $T=1$ and set $d_{\max}=4T$.
We consider $n_b=4$ fading blocks. 
Oversampling by $N=8$ gives $t_s=T/8$; so each block contains $M=32+128=160$ samples. 
As in \cite{dornerLearningJointDetection2022}, we use Zadoff--Chu pilots~\cite{TS138211} for the 5G baselines and the \ac{RCUs} bound, which is optimized over $s$ and $n_p$.
The 5G baselines are simulated with Sionna~\cite{Sionna}. 
Under block fading, the optimized 5G baselines (Section~\ref{sec:5G_baseline}) involve the transmission of $n_p=2$ pilots per block, $16$-PSK for 5G polar + CRC SC, and $32$-PSK for both 5G polar + CRC SCL-8 and 5G LDPC BP-20. 
The \ac{TurboAE} is trained with Adam for $700$ steps with an initial batch size of $500$ and a learning rate of $10^{-4}$.

To single out the benefit of joint synchronization, channel estimation, and decoding, we first consider an \ac{AWGN} channel with perfect synchronization, where the baselines use QPSK. In this case (Fig.~\ref{fig:ber_awgn}), TurboAE has the lowest BER up to about $\rho = \qty{3.3}{\decibel}$, but its BER decays more slowly than that of 5G polar + CRC SCL-8 and 5G LDPC BP-20. Hence, it offers no coding gain over them at moderate-to-high SNR. 

For the considered block-fading channel (Fig.~\ref{fig:ber_fading}), instead, TurboAE achieves the lowest BER over the entire considered SNR range and requires \qty{1.9}{\decibel} less SNR than 5G LDPC BP-20 to achieve a BER of $10^{-3}$. We attribute this advantage to joint synchronization and decoding, which avoids the pilot overhead. This is consistent with~\cite[Fig.~13]{dornerLearningJointDetection2022}, where, for the same packet size and rate, the TurboAE-based scheme outperforms a LDPC baseline with a 20-symbol preamble by \qty{2.7}{\decibel} in BER over a multipath channel.

In terms of \ac{BLER} under AWGN (Fig.~\ref{fig:bler_awgn}), TurboAE requires a higher SNR than 5G polar + CRC SCL-8, 5G LDPC BP-20, and the \ac{RCUs} bound. Instead, for the considered block-fading channel (Fig.~\ref{fig:bler_fading}), TurboAE slightly outperforms 5G polar + CRC SC but, at a BLER of $10^{-3}$, requires $1.4$, $2.4$, and \qty{4.5}{\decibel} more SNR than 5G LDPC BP-20, 5G polar + CRC SCL-8, and the \ac{RCUs} bound, respectively. Hence, the \qty{1.9}{\decibel} BER gain over 5G LDPC BP-20 turns into a \qty{1.4}{\decibel} BLER loss. As also noted in~\cite{dornerLearningJointDetection2022}, this discrepancy between BER and BLER performance may be partly due to the use of the \ac{BCE} loss~\eqref{eq:turboae_bit_loss}, which targets individual bits rather than whole packets. Inspired by~\cite{dornerLearningJointDetection2022}, we consider a concatenated scheme, in which we (i) append a 3-bit \ac{CRC} with generator polynomial $g(x)=x^3+x+1$, (ii) retain at the decoder the eight most likely candidates under the predicted bit probabilities, and (iii) select the most likely CRC-valid candidate. This improves the TurboAE BLER by about \qty{1.6}{\decibel}. On the contrary, finetuning with a BLER-targeted \emph{LogSumExp} loss~\cite{wiesmayrBitErrorBlock2023a} (see Fig.~\ref{fig:bler_fading}) or adding a block-reconstruction penalty to the loss~\eqref{eq:turboae_bit_loss} improves the BLER only slightly.

\section{Conclusion}
We developed an end-to-end learned scheme for short-packet transmission over a continuous-time waveform channel with an unknown delay and memoryless block fading. In the proposed scheme, the delay is estimated without dedicated pilots. The scheme contains an encoder, a synchronizer, and a decoder, which are trained alternately. Under the same blocklength and rate constraints, TurboAE achieves a lower BER than all evaluated pilot-assisted 5G baselines but a higher BLER than 5G polar + CRC SCL-8; with a 3-bit \ac{CRC}, its BLER is comparable to that of 5G LDPC BP-20. Finally, our work uncovered a large BLER gap between all evaluated schemes and the RCUs bound, motivating future work on new architectures and training methods for short-packet communication. 

\bibliographystyle{IEEEtran}
\bibliography{references/references}

\end{document}